%% file: main.tex
\documentclass[conference,10pt]{IEEEtran}
\IEEEoverridecommandlockouts

\usepackage{amsmath,amsfonts}
\usepackage{graphicx}
\usepackage{cite}
\usepackage{amssymb}
\usepackage{booktabs}
\usepackage[dvipsnames]{xcolor}
\usepackage{enumerate}
\usepackage{pgfplots}
\usepackage{mathtools}
\usepackage{pstool}
\usepackage{psfrag}
\usepackage[]{algpseudocode}
\newcommand{\algmargin}{\the\ALG@thistlm}
\usepackage[ruled,vlined,linesnumbered,lined,commentsnumbered]{algorithm2e}
\usepackage[printonlyused,withpage]{acronym}
\usepackage{caption}
\usepackage{subcaption}
\usepackage[table]{xcolor}
\usepackage[bottom]{footmisc}
\pgfplotsset{compat=1.18}
\input{Acronyms.tex}  

\begin{document}

\title{Dual-Chirp AFDM for Joint Delay-Doppler \\ \!Estimation with Rydberg Atomic Quantum Receivers\!
\thanks{The fundamental research described in this paper was supported by the National Research Foundation (NRF) of Korea under Grant RS-2024-00409492, and by the German Research Foundation (DFG) through the QUBYSM Project with Grant No. G:(GEPRIS)576171458.}}
\author{
\IEEEauthorblockN{
Hanvit Kim\IEEEauthorrefmark{1}, 
Hyeon Seok Rou\IEEEauthorrefmark{2}, Kihong Min\IEEEauthorrefmark{1},
Giuseppe Thadeu Freitas de Abreu\IEEEauthorrefmark{2}, and Sunwoo Kim\IEEEauthorrefmark{1} \\[1ex]
}
\IEEEauthorblockA{\IEEEauthorrefmark{1}Department of Electronic Engineering, Hanyang University, Republic of Korea}
\IEEEauthorblockA{\IEEEauthorrefmark{2}School of Computer Science and Engineering, Constructor University Bremen, Germany \\[1ex]}
\IEEEauthorblockA{\small Emails: \{dante0813, khmin705, remero\}@hanyang.ac.kr, \{hrou, gabreu\}@constructor.university} \vspace{-5ex}
}

\maketitle

\begin{abstract}
In this paper, we propose a joint delay-Doppler estimation framework for \ac{RAQRs} leveraging \ac{AFDM}, as a future enabler of hyper \ac{ISAC} in 6G and beyond.
The proposed approach preserves the extreme sensitivity of \ac{RAQRs}, while offering a pioneering solution to the joint estimation of delay-Doppler parameters of mobile targets, which has yet to be addressed in the literature due to the inherent coupling of time-frequency parameters in the optical readout of \ac{RAQRs} to the best of our knowledge.
To overcome this unavoidable ambiguity, we propose a dual-chirp \ac{AFDM} framework where the utilization of distinct chirp parameters effectively converts the otherwise ambiguous estimation problem into a full-rank system, enabling unique delay-Doppler parameter extraction from \ac{RAQRs}.
Numerical simulations verify that the proposed dual-chirp AFDM shows superior delay-Doppler estimation performance compared to the classical single-chirp \ac{AFDM} over \ac{RAQRs}.
\end{abstract}

\begin{IEEEkeywords}
Quantum sensing, Rydberg atomic quantum receivers (RAQRs), delay-Doppler estimation, affine frequency division multiplexing (AFDM), ISAC.
\end{IEEEkeywords}

\acresetall	

\section{Introduction}
\label{sec:introduction}

Next generation \ac{ISAC} for 6G and beyond requires extremely high-quality wireless connectivity and robust sensing capability in the high-mobility scenario \cite{liu2022integrated}.
The high-mobility scenario results in a doubly dispersive wireless channel, posing a significant challenge for accurate channel parameter estimations, such as propagation delay and Doppler shifts \cite{Rou_SPM24}.
To solve this problem, \ac{AFDM} waveform has recently been proposed, which enables accurate delay-Doppler estimation by reducing the \ac{ICI} arising from the severe Doppler shifts in doubly-dispersive channel \cite{bemani2023affine}.
Furthermore, its significant implications for standardization highlight its potential as a leading candidate for next-generation wireless systems \cite{Rou_CSM26}.
Indeed, the \ac{AFDM} have shown advantages in both sensing and communications functionalities in doubly-dispersive channels compared to the other emerging waveforms for \ac{ISAC}, such as \ac{OTFS}\cite{ranasinghe2024joint}.

Aside from the waveform design, recent studies are focusing on the \ac{RAQRs}, which are envisioned to be a key enabler for high-sensitivity wireless sensing \cite{gong2025rydberg}.
Harnessing the extreme sensitivity of Rydberg atoms, \ac{RAQRs} can approach the \ac{SQL} ($\sim 700 \mathrm{pV} \cdot \mathrm{~cm}^{-1} \cdot \mathrm{~Hz}^{-1 / 2}$) of electric-field sensitivity, which is significantly lower than the thermal noise of conventional dipole antenna-based receivers \cite{jing2020atomic}.
Recently, \ac{RAQRs} have facilitated significant advancements across various sensing applications, including \ac{AOA} estimation \cite{kim2025quantum}, Doppler-based localization \cite{guo2025ultra},  and \ac{ISAC} \cite{chen2025new}. 

Despite recent advancements, utilizing \ac{RAQRs} for precise delay-Doppler estimation remains an unresolved challenge.
This is primarily due to two factors: \textit{1)} the absence of a system model and framework for joint delay-Doppler estimation with \ac{AFDM} using \ac{RAQRs}, and \textit{2)} the intrinsic delay-Doppler ambiguity introduced by the optical readout of Rydberg probes in doubly-dispersive channels.
While several \ac{RAQRs} system models have been proposed \cite{cui_MIMO, gong2024rydberg}, these consider a single-carrier waveform, which are not applicable for the multi-carrier \ac{AFDM}-based delay-Doppler estimation.
Some initial attempts include range estimation with \ac{RAQRs} \cite{cui2025realizing}, but joint delay-Doppler estimation for 6G \ac{ISAC} applications still remains an unaddressed issue.

In this article, we address these limitations by proposing a delay-Doppler estimation framework for \ac{RAQRs} utilizing \ac{AFDM} waveforms. 
By deriving the \ac{RF}-to-optical measurement model for \ac{AFDM} signals, we employ the atomic autocorrelation method \cite{cui2025realizing} to map delay-Doppler parameters onto the frequency domain. 
This analysis reveals that conventional single-chirp \ac{AFDM} signals yield ambiguous measurements, significantly complicating the joint estimation process. 
To resolve this, we introduce a dual-chirp \ac{AFDM} variant that leverages two distinct post-chirp rates, $c^{(A)}_{1}$ and $c^{(B)}_{1}$, which facilitates rank-restoration of the estimation model from the optical measurements, enabling unambiguous estimation of the delay and Doppler parameters of the radar targets.

The contributions of the article are summarized as follows: \vspace{-3ex}
\begin{itemize}
\item The quantum system model for \ac{AFDM} signal detection is analyzed, and a dedicated framework for joint delay-Doppler estimation utilizing \ac{RAQRs} is established;
\item The dual-chirp \ac{AFDM} is proposed to resolve the rank-deficient joint delay-Doppler estimation problem, and an \ac{LS}-based estimation algorithm is proposed to achieve high-accuracy parameter recovery.
\end{itemize}


\section{System Model}
\label{sec:system_model}
\subsection{Affine Frequency Division Multiplexing}
The transmitted \ac{DAFT}-domian vector $x[m]$ is mapped onto the discrete time domain signal $s[n]$ using the \ac{IDAFT}, which is presented as \cite{AFDM_Marios}
\begin{equation}
s[n]=\frac{1}{\sqrt{N}} \sum_{m=0}^{N-1} x[m] e^{j 2 \pi\left(c_2 m^2+\frac{1}{N} m n+c_1 n^2\right)},\label{tx_sig_discrete}
\end{equation}
where $N$ is the number of chirp subcarriers, and $c_{1}$ and $c_{2}$ are the pre-chirp and post-chirp parameters of the \ac{IDAFT}, affecting various properties of \ac{AFDM}.

Given the above, the continuous time version of the transmitted signal in \eqref{tx_sig_discrete} can be written as \cite{yin2025ambiguity}
\begin{equation}
s(t)= \sum_{m=0}^{N-1} x[m] e^{j 2 \pi(c_{2}m^2+\phi_{m}(t))}, \quad 0 \leq t \leq T.
\label{tx_sig}
\end{equation}
where $T=N \Delta t$, with the Nyquist sampling rate $\frac{1}{\Delta t}$, is the duration of the instantaneous phase function of the $m$-th chirp $\phi_{m}(t)$, defined as piece-wise manner as \cite{AFDM_Marios}
\begin{equation} \label{instantaneous_phase}
\phi_m(t)=\tilde{c}_1 t^2+\frac{m}{T}t-\frac{q}{\Delta t}t, \quad t_{m, q} \leq t<t_{m, q+1},
\end{equation}
with $\tilde{c}_{1}=\frac{c_{1}}{(\Delta t)^2}$, and $t_{m,q}$ is the $q$-th spectrum wrapping point of $m$-th chirp subcarrier.
Note that $c_1$ controls the frequency dispersion of the signal, which can be changed and optimized according to the Doppler characteristics of the doubly dispersive wireless channel \cite{tek2025novel}.

\subsection{Rydberg Atomic Quantum Receiver Model}
\subsubsection{Quantum state} 

The energy level of an atom changes as the photon is either absorbed or emitted. 
By leveraging this electron transition, the \ac{RF} signals can be detected. 
The electron transition can be modeled by different quantum states of the Rydberg atom. 
These quantum states include the ground state $|1\rangle$, a lowly-excited state $|2\rangle$, and the Rydberg states, which are $|3\rangle$ and $|4\rangle$.
The probe beam of angular frequency $\omega_{p}$ excites the quantum state from $|1\rangle \rightarrow |2\rangle$, and the coupling beam of angular frequency $\omega_{c}$ induces the transition $|2\rangle \rightarrow |3\rangle$, transforming the alkali-metal atom to the Rydberg atom.
Thereafter, the RF signals of angular frequency $\omega_{\mathrm{RF}}$ excite the Rydberg state to another Rydberg state $|3\rangle \rightarrow |4\rangle$, enabling the \ac{RF} signal detection by monitoring the variations induced by these electron transitions via \ac{PD}.

\subsubsection{Rabi frequency and detuning} 

The interaction strength between the \ac{RF} signals and the electric dipole moment is characterized by the \textit{Rabi frequency}. 
The general expression of Rabi frequency $\Omega_{\mathrm{RF}}$ is given by \cite{jing2020atomic}
\begin{equation}
\Omega_{\mathrm{RF}}=\frac{\mu_{34}}{\hbar}|E_{\mathrm{RF}}|,
\label{General_Rabi}
\end{equation}
where $\mu_{34}$, $\hbar$, and $E_{\mathrm{RF}}$ are the transition dipole moment, reduced Planck constant, and RF signals, respectively. 

Furthermore, the frequency deviation between the transition frequency and the carrier frequency arises for every electron transition process due to the discrete energy level, also referred to as \textit{frequency detuning}. 
For example, the frequency detuning $\Delta_{p}$ presents the gap between $\omega_{p}$ and the transition frequency $\omega_{12}$, i.e., $\Delta_{p}=\omega_{p}-\omega_{12}$.
Likewise, the detuning of the coupling beam $\Delta_{c}$ and the RF signal $\Delta_{\mathrm{RF}}$ can be introduced in the same manner so that $\Delta_{c}=\omega_{c}-\omega_{23}$ and $\Delta_{\mathrm{RF}}=\omega_{\mathrm{RF}}-\omega_{34}$, where $\omega_{23}$ and $\omega_{34}$ are transition frequency of coupling beam and RF signals, respectively.
The detunings are zero when they are on-resonant with their respective electron transitions. In this work, we assumed the probe beam and the coupling beam are on-resonant with their transitions (i.e., $\Delta_{p}=\Delta_{c}=0$) \cite{cui2025realizing}.

\subsubsection{Dynamics of quantum state}
The dynamic quantum state is governed by the Lindblad master equation.
For the \ac{AFDM}, we adopt the four-level system model since the deviation of each chirp subcarrier frequency from the transition frequency can be fully described by the different frequency detunings.
Based on this system, the Lindblad master equation is represented as \cite{jing2020atomic}
\begin{equation}
\frac{\partial \boldsymbol{\rho}}{\partial t}=-\frac{j}{\hbar}[\mathbf{H}, \boldsymbol{\rho}]+\mathcal{L},
\label{Lindblad}
\end{equation}
where $[\mathbf{H}, \boldsymbol{\rho}]$ denotes the commutator $\mathbf{H}\boldsymbol{\rho}-\boldsymbol{\rho}\mathbf{H}$.
Here, $\mathbf{H}$, $\boldsymbol{\rho}$, and $\mathcal{L}$ are the Hamiltonian operator, density matrix, and the decoherence operator, respectively. 
Here, the Hamiltonian operator $\mathbf{H}$ is given by
\begin{equation}
\mathbf{H}=\frac{\hbar}{2}\left[\begin{array}{cccc}
0 & \Omega_{p} & 0 & 0 \\
\Omega_{p} & 0 & \Omega_{c} & 0 \\
0 & \Omega_{c} & 0 & \Omega_{\mathrm{RF}} \\
0 & 0 & \Omega_{\mathrm{RF}} & -2 \Delta_{\mathrm{RF}}
\end{array}\right],
\label{Hamiltonian}
\end{equation}
where $\Omega_{p}$, $\Omega_{c}$, $\Omega_{\mathrm{RF}}$, and $\Delta_{\mathrm{RF}}$ are Rabi frequencies of probe beam, coupling beam, \ac{RF} signal, and the frequency detunings of \ac{RF} signal, respectively. 
The decoherence matrix $\mathcal{L}$ is presented as \cite{jing2020atomic}
\begin{equation}
\mathcal{L}=-\frac{1}{2}\{\boldsymbol{\Gamma}, \boldsymbol{\rho}\}+\boldsymbol{\Lambda},
\label{Decay_Matrix}
\end{equation}
where $\{\boldsymbol{\Gamma}, \boldsymbol{\rho}\}$ denotes the anti-commutator $\boldsymbol{\Gamma}\boldsymbol{\rho}+\boldsymbol{\rho}\boldsymbol{\Gamma}$ and $\boldsymbol{\Gamma}=\operatorname{diag}\left\{0, \gamma_2, \gamma_3, \gamma_4\right\}$. Here, $\gamma_{j}$ are the decay rates of the $j$-th level.
The decay matrix is presented as $\boldsymbol{\Lambda}=\operatorname{diag}\left\{\gamma_2 \rho_{22}+\gamma_4 \rho_{44}, \gamma_3 \rho_{33}, 0,0\right\}$.

\subsubsection{Measured optical signal}
The optical measurement, which corresponds to the probe beam output power, can be acquired by solving the steady-state solution of $\rho_{12}$, which is the $(1,2)$-th element of the density matrix $\boldsymbol{\rho}$
Here, the $\rho_{12}$ is given by \cite{cui2025realizing}
\begin{equation}
\rho_{12}=\frac{A_{1} \Omega_{\mathrm{RF}}^2 \Delta_{\mathrm{RF}}^2+j B_1 \Omega_{\mathrm{RF}}^4}{C_1 \Omega_{\mathrm{RF}}^4+C_2 \Omega_{\mathrm{RF}}^2+C_3 \Delta_{\mathrm{RF}}^2},
\label{rho_12}
\end{equation}
where $A_1=2 \Omega_p \Omega_c^2$, $B_1=\gamma_2 \Omega_p$, $C_1=2 \Omega_p^2+\gamma_2^2$, $C_{2}=2 \Omega_p^2\left(\Omega_c^2+\Omega_p^2\right)$, and $C_3=4\left(\Omega_c^2+\Omega_p^2\right)^2$. 

Let $P_{\mathrm{in}}$ represents the input power of the probe beam. Then, according to the adiabatic approximation, the output power of the probe beam $P_{\mathrm{out}}$ is defined by the imaginary part of the $(1,2)$-th entry of $\boldsymbol{\rho}$, $\rho_{12}$, which is presented as \cite{jing2020atomic}
\begin{equation}
P_{\mathrm{out}}=  P_{\text {in}} \exp \left(-C_{0} \operatorname{Im}\left\{\rho_{12}\right\}\right),
\label{Output_Power}
\end{equation}
where the constant $C_{0}$ is given by $C_{0} \triangleq \frac{\Delta2 N_0 \mu_{12}^2 k_p L}{\epsilon_0 \hbar \Omega_{p}}$.
Here, $N_{0}$, $\mu_{12}$, ${\epsilon}_{0}$, $L$, and $k_{p}=\frac{2 \pi}{\lambda_{p}}$ are density of atoms, transition dipole moment of transition $|1\rangle \rightarrow |2\rangle$, vacuum permittivity, length of vapor cell, and wavenumber of probe beam with probe beam wavelength $\lambda_{p}$, respectively.

After the readout of the probe beam power, the \ac{PD} converts it into the current $I_{\mathrm{out}}=\frac{q \eta}{\hbar \omega_{\mathrm{p}}} P_{\mathrm{out}}$, where $\eta$ and $q$ are the quantum efficiency of the \ac{PD} and the charge of electrons.
Then, the output voltage is given by \cite{cui2025realizing}
\begin{equation}
V_{\mathrm {out }}=R_{\mathrm{T}} I_{\mathrm {out }} \triangleq V_{\mathrm {in }} \exp \left(-C_0 \operatorname{Im}\left\{\rho_{12}\right\}\right),
\label{Output_Voltage}
\end{equation}
where $R_{\mathrm{T}}$ is the load impedance and $V_{\mathrm{in}} \triangleq \frac{R_{\mathrm{T}} q \eta}{\hbar \omega_{\mathrm{p}}} P_{\mathrm{in}}$ is the input voltage.

We define the bias function $\Pi(\Omega, \Delta)$ of the probe beam output power by substituting \eqref{rho_12} into \eqref{Output_Power}, as
\begin{equation}
\Pi(\Omega, \Delta) \triangleq V_{\mathrm{in }} \exp \left\{-\frac{B_1 C_0 \Omega^4}{C_1 \Omega^4+C_2 \Omega^2+C_3 \Delta^2}\right\},
\label{bias}
\end{equation}
where $\Omega \in[0,+\infty)$ and $\Delta \in \mathbb{R}$ are the general notation of the Rabi frequency and detuning, respectively.

Furthermore, the gain of the probe beam power $\Upsilon(\Omega, \Delta)$ is derived as a partial derivative of $\Pi(\Omega, \Delta)$, so that $\Upsilon(\Omega, \Delta)\triangleq \frac{\partial \Pi(\Omega, \Delta)}{\partial \Omega}$, where detailed presentation can be found in \cite{cui2025realizing}.
%

\section{Problem Formulation}
\label{sec:problem_formulation}
\subsection{Measured \ac{AFDM} Signal Model}

\begin{figure}[t]
\centering
\includegraphics[width=\linewidth]{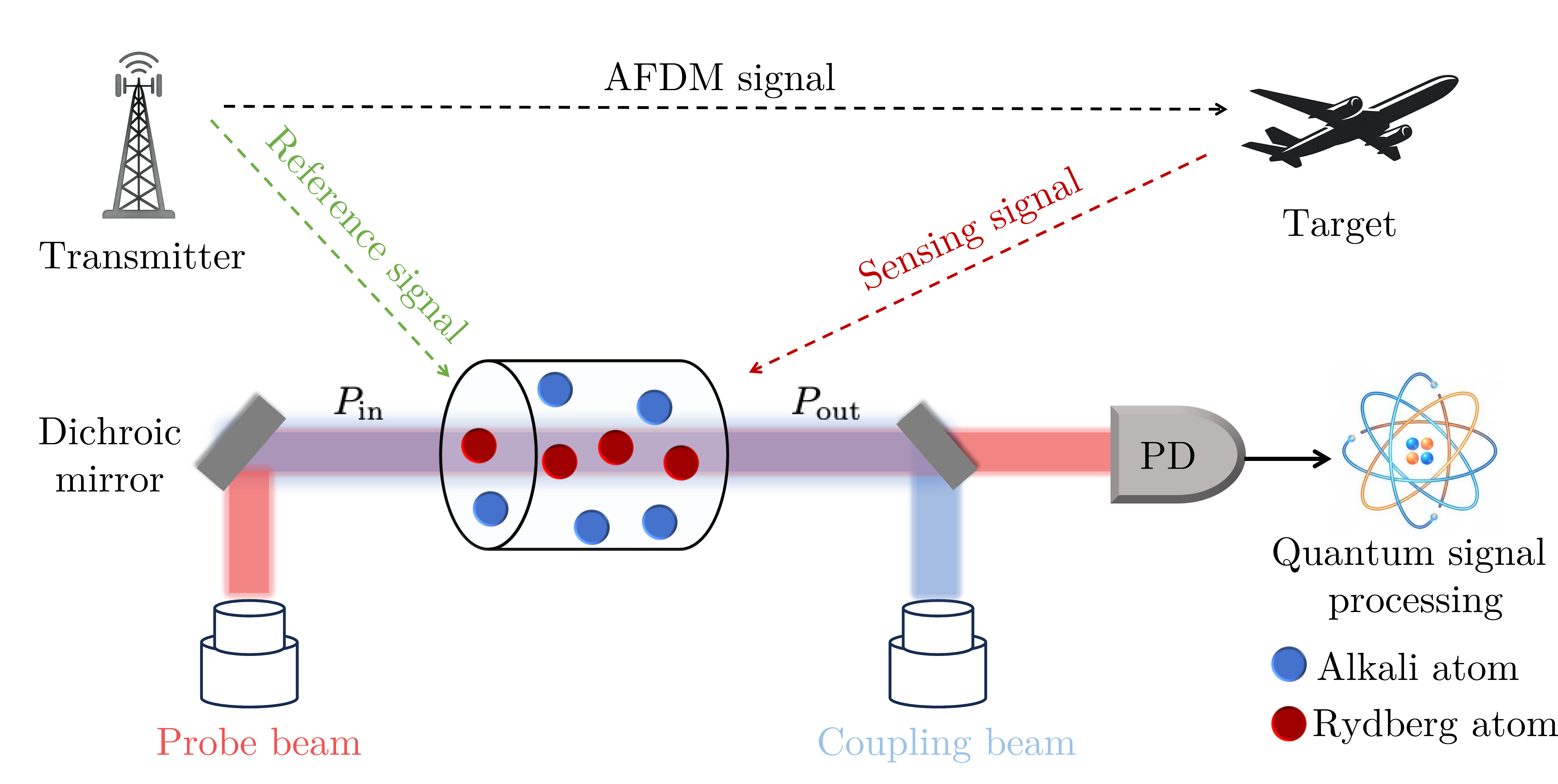}
\caption{Quasi-monostatic sensing scenario with AFDM signal for delay-Doppler estimation with \ac{RAQRs}.}
\label{Scenario}
\vspace{-2ex}
\end{figure}

Consider a quasi-monostatic sensing of a single moving target in the high-mobility scenario, resulting in a doubly dispersive wireless channel \cite{Rou_SPM24}.
As illustrated in Fig.~\ref{Scenario}, we assume the transmitter broadcasts the \ac{AFDM} signal in downlink to a single moving target and the static \ac{RAQR}.
Following the received AFDM-\ac{ISAC} signal model in \cite{yin2025ambiguity}, the Rabi frequency of the reference signal $\Omega_{l}$ (line-of-sight only) is
\begin{equation}
\Omega_{l}=\frac{\mu_{34}}{\hbar}\left|h_{l}s(t-\tau^{\prime})e^{j 2 \pi \nu_l t} \right|,
\label{Rabi_Ref}
\end{equation}
where $h_{l}, \,\nu_{l}$, and $\tau^{\prime}$ are the channel gain, Doppler shift, and the propagation delay of the \ac{LOS} path between transmitter and \ac{RAQR}, respectively. 
Trivially, $\nu_{l}=0$ since the \ac{RAQR} is fixed without mobility.

We employ the self-heterodyne sensing technique \cite{cui2025realizing}, in which the transmitter itself act as the \ac{LO} to extract the delay-Doppler from the received signal. 
In this scheme, the frequency of the measured signal is presented as the phase function between the phase profile difference of \ac{LO} and the target.
Based on \eqref{instantaneous_phase}, the phase function for \ac{AFDM} $\Delta\phi_{m}(t,\tau,\tau^{\prime},\nu_{r})$ in a doubly-dispersive channel is given by
\begin{align} \label{phase_diff_AFDM}
\Delta \phi_m\left(t, \tau, \tau^{\prime},\nu_{r}\right) & =\phi_m\left(t-\tau\right)-\phi_m\left(t-\tau^{\prime}\right)+\nu_{r}t  \\
& =\left(2 \tilde{c}_1\left(\tau^{\prime}-\tau\right)+\nu_{r}\right) t  \nonumber \\ &+\left(\tau-\tau^{\prime}\right)\left(\frac{m}{T}-\frac{q}{\Delta t}+\tilde{c}_1(\tau+\tau^{\prime})\right), \nonumber
\end{align}
where $\nu_{r}=\frac{2v_{r}\omega_{\mathrm{RF}}}{2 \pi c}$ is the Doppler shift of the target, with the relative radial velocity $v_{r}$ and speed of light $c$, and the propagation delay of the target-to-\ac{RAQR} path $\tau$.

Furthermore, the instantaneous frequency of \ac{AFDM} coincides with the derivative of $\phi_{m}(t)$ \cite{AFDM_Marios}. 
Therefore, the frequency detuning of the reference signal at the $m$-th chirp subcarrier $\Delta_{l,m}(t)$ is presented as
\begin{equation} \label{detuning_ref}
\begin{aligned}
\Delta_{l,m}(t) & =\frac{d \phi_{m}(t-\tau^{\prime})}{dt}-\omega_{34} \\ & = 2\tilde{c}_{1}t-2\tilde{c}_{1}\tau^{\prime}+\frac{m}{T}-\frac{q}{\Delta t}-\omega_{34}.
\end{aligned}
\end{equation}
The noise of the $m$-th chirp $n_{m}(t)$ is modeled as Gaussian noise $n_{m}(t)\sim\mathcal{CN}(0,\sigma^2_{m})$, and the intrinsic noise and the extrinsic noise are presented as
\begin{align}\small\label{noise_intrinsic_2}
\sigma_{\mathrm{Int},m}^2(t)&=q R_{\mathrm{T}} \Pi\left(\Omega_l, \Delta_{l,m}(t)\right), \notag \\ 
\sigma_{\mathrm{Ext},m}^2(t)&=\frac{\mu_{34}^2}{\hbar^2} \Upsilon^2\left(\Omega_l, \Delta_{l,m}(t)\right)\left\langle E_I^2\right\rangle, 
\end{align}
where $\left\langle E_I^2\right\rangle=\frac{\hbar \omega_{\mathrm{RF}}^3}{\pi \epsilon_0 c^3}\left(2 n_{\mathrm{th}}+1\right)$ is the field intensity of blackbody radiation \cite{cui2025realizing}.
Here, $\epsilon_0$ is the dielectic constant and $n_{\mathrm{th}}=1 /\left(e^{\hbar \omega_{\mathrm{RF}} / k_B T_E}-1\right)$ is the Bose-Einstein distribution, where $k_{B}$ and $T_{E}$ are Boltzmann constant and the ambient temperature, respectively.

Given the above, the noise power $\sigma_{m}^2(t)$ is given by
\begin{equation} \label{noise_tot}
\sigma_{m}^2(t)=\sigma_{\mathrm{Int},m}^2(t)+\sigma_{\mathrm{Ext},m}^2(t),
\end{equation}
and consequently, the measured \ac{AFDM} signal at the $m$-th chirp subcarrier is presented as
\begin{align}\label{power_AFDM}
y_{m}(t)= & \;\Pi\left(\Omega_l, \Delta_{l, m}(t)\right)+\frac{\mu_{34}}{\hbar} \Upsilon\left(\Omega_l, \Delta_{l, m}(t)\right) \nonumber \\[-0.5ex]
& ~~\times \frac{1}{N} \sqrt{P_s} \cos \left(\Delta \phi_m\left(t , \tau, \tau^{\prime}, \nu_r\right)\right)+n_m(t),
\end{align}
where $P_{s}$ denotes the power of the received signal.

\subsection{Delay-Doppler Parameter Estimation Problem}\label{sec3-2}

In light of the measured signal model formulation of \ac{AFDM} in doubly dispersive channel under \ac{RAQRs}, the objective of this paper is to accurately estimation the delay-Doppler information of the target $[\tau,\nu_{r}]$ from the measurements.
To achieve this goal, we leverage the atomic autocorrelation property of self-heterodyne sensing, which maps the target's delay to a fluctuation frequency of the output voltage of \ac{RAQRs} \cite{cui2025realizing}.

Mathematically, the instantaneous phase function in \eqref{phase_diff_AFDM} is modeled as the fluctuation frequency and the phase of the output voltage, which is presented as
\begin{equation} \label{phase_re}
\Delta \phi_m\left(t, \tau, \tau^{\prime}, \nu_r\right)=-\omega_{m}t-\varphi_{m},
\end{equation}
where $\omega_{m}$ and $\varphi_{m}$ are fluctuation frequency and phase of the $y_{m}(t)$.
Then, by substituting \eqref{phase_re} into the \eqref{phase_diff_AFDM}, the observed fluctuation frequency $\omega_{m}$ is given by
\begin{equation} \label{probe_freq}
\omega_{m}=2 \tilde{c}_1\left(\tau-\tau^{\prime}\right)-\nu_r.
\end{equation}
From \eqref{probe_freq}, one can observe that the two key parameters to be estimated -- time delay $\tau$ and the Doppler shift $\nu_{r}$ of the target -- are correspondingly embedded in the fluctuation frequency $\omega_{m}$.
In other words, this atomic correlation property of self-heterodyne sensing converts the joint delay-Doppler estimation into the fluctuation frequency estimation problem.

For $N$ subcarriers, we define the frequency vector $\boldsymbol{\omega}$ as
\begin{equation} \label{Measured_Freq_Vec}
\boldsymbol{\omega}=\left[\begin{array}{c}
\omega_0 \\
\vdots \\
\omega_{N-1}
\end{array}\right]=\left[\begin{array}{c}
2 \tilde{c}_1 \tau-2 \tilde{c}_1 \tau^{\prime}-\nu_r \\
\vdots \\
2 \tilde{c}_1 \tau-2 \tilde{c}_1 \tau^{\prime}-\nu_r
\end{array}\right],
\end{equation}
which easily confirms that the estimation problem of $\tau$ and $\nu_r$ from the fluctuation frequencies $\{\omega_0,\ldots,w_{N-1}\}$ is an ill-posed linear inverse problem since there are two unknown target parameters, $\tau$ and $\nu_{r}$, with $N$ equivalent equations (i.e., $\omega_{0}=\omega_{1}=\cdots=\omega_{N-1}$).
Therefore, with the conventional \ac{AFDM}, it is highly challenging to extract the delay-Doppler information due to the under-determined system model, which motivates the design of a modified \ac{AFDM} waveform for the \ac{RAQR}-enabled joint delay-Doppler estimation. 

\section{Proposed Dual-Chirp AFDM-enabled \\ Joint Delay-Doppler Estimator}
\label{sec:proposed}
\subsection{Dual-Chirp AFDM}\label{sec4-1}

To estimate the delay-Doppler from the optical measurement of \ac{RAQRs}, we propose a modified \ac{AFDM} signal, termed as the dual-chirp AFDM, which utilizes two distinct post-chirp rates consecutively, denoted as $c^{(A)}_{1}$ and $c^{(B)}_{1}$ respectively.
Then, the transmitted signal with a new frame structure of dual-chirp \ac{AFDM} is described by
\begin{equation} 
s(t)= \begin{cases} \sum_{m=0}^{N-1} x[m] e^{j 2 \pi(c_{2}m^2+\phi^{(A)}_{m}(t))}, & 0 < t \leq {T}^{\prime} \\  \sum_{m=0}^{N-1} x[m] e^{j 2 \pi(c_{2}m^2+\phi^{(B)}_{m}(t))}, & T^{\prime} < t \leq T .\end{cases}
\label{DC-AFDM_Frame}
\end{equation}
where $T^{\prime}$ is the duration of phase function $\phi^{(A)}_{m}(t)$ for post-chirp $c^{(A)}_{1}$.
Note that the number of chirp subcarriers for $\phi^{(B)}_{m}(t)$ is equal to that of $\phi^{(A)}_{m}(t)$, and the phase function $\phi^{(A)}_{m}(t)$ and $\phi^{(B)}_{m}(t)$ of dual-chirp AFDM are given by
\begin{align}\small\label{Phase-DC-AFDM}
\phi^{(A)}_m(t)&=\tilde{c}^{(A)}_1 t^2+\frac{m}{T^{\prime}}t-\frac{q}{\Delta t}t, \notag \\ 
\phi^{(B)}_m(t)&=\tilde{c}^{(B)}_1 t^2+\frac{m}{T-T^{\prime}}t-\frac{q}{\Delta t}t, 
\end{align}
where $\tilde{c}^{(A)}_1=\frac{c^{(A)}_{1}}{\Delta t^2}$ and $\tilde{c}^{(B)}_1=\frac{c^{(B)}_{1}}{\Delta t^2}$, respectively.

Then, by substituting \eqref{Phase-DC-AFDM} into \eqref{phase_diff_AFDM}, the fluctuation frequency of the $m$-th subcarrier in \eqref{probe_freq} is reformulated as
\begin{equation} 
\!\!\!\omega_{m}= \begin{cases} 2 \tilde{c}^{(A)}_1(\tau-\tau^{\prime})-\nu_r, & 0 < t \leq {T}^{\prime} \\  2 \tilde{c}^{(B)}_1( \tau-\tau^{\prime})-\nu_r, & T^{\prime} < t \leq {T}.\end{cases} 
\label{DC-AFDM_Re} \vspace{-1.5ex}
\end{equation}

From equation \eqref{DC-AFDM_Re}, it is trivial that the problem is now solvable since the utilization of the two post-chirps, $c^{(A)}_{1}$ and $c^{(B)}_{1}$, introduces an additional distinct equation to the system.

\subsection{Proposed Algorithm}\label{sec4-2}
%
%
%

\subsubsection{Fluctuation frequency estimation} We first estimate the fluctuation frequency of the measured signal.
Here, we adopt the \ac{NLS}-based optimization algorithm in \cite{cui2025realizing}.
First, the normalized received signal of the $m$-th chirp subcarrier is presented as 
\begin{align} \label{norm_received}
\bar{y}_{m}(t) & =\frac{y_{m}(t)-\Pi\left(\Omega_l(t), \Delta_{l,m}(t)\right)} {\sigma_{m}(t)} \notag \\
& =\varrho_{m}(t) h \cos (\omega_{m} t+\varphi_{m})+\bar{n}_{m}(t),
\end{align}
where $\bar{n}_{m}(t)$ is a unit power of Gaussian noise and $h$ denotes a gain of \ac{LOS} channel.
Here, the amplitude $\varrho_{m}(t)$ is given by
\begin{equation} \label{norm_gain}
\varrho_{m}(t)=\frac{\mu_{34} \Upsilon\left(\Omega_{l}, \Delta_{l,m}(t)\right) \sqrt{P_{s}}}{\hbar \sqrt{N(\sigma_{\mathrm{Int},m}^2(t)+\sigma_{\mathrm{Ext},m}^2(t)})} .
\end{equation}
To ease the expression, consider a time duration for $\phi_{m}^{(A)}(t)$. Then, following the \ac{NLS} problem in \cite{cui2025realizing}, the frequency estimation problem is presented as 
\begin{equation} \label{NLS_Prob}
\max _{{\omega}^{(0)}_{m}}\left|\int_0^{T^{\prime}} \bar{y}_{m}(t) \varrho_{m}(t) e^{j \omega^{(0)}_{m} t} \mathrm{~d} t\right|^2.
\end{equation}
The peaks of $\left|\int_0^{T^{\prime}} \bar{y}_{m}(t) \varrho_{m}(t) e^{j \omega^{(0)}_{m} t} \mathrm{~d} t\right|^2$ can be observed via time-frequency analysis technique such as \ac{FFT}, where its location $\boldsymbol{\omega}^{(0)}=[\omega_{0}^{(0)},\cdots,\omega_{N-1}^{(0)}]^{\mathsf{T}}$ represents the estimated frequency for all chirp subcarriers.

After the initial estimation, the refinement is conducted to enhance the accuracy.
By using Newton's method, the refinement is presented as
\begin{equation} \label{Opt}
\binom{\hat{\omega}_{m}^{(t)}}{\hat{\varphi}_{m}^{(t)}}=\binom{\hat{\omega}_{m}^{(t-1)}}{\hat{\varphi}_{m}^{(t-1)}}+\left(\begin{array}{cc}
\frac{\partial^2 Q}{\partial \hat{\omega}_{m}^2} & \frac{\partial^2 Q}{\partial \hat{\omega}_{m} \partial \hat{\varphi}_{m}} \\
\frac{\partial^2 Q}{\partial \hat{\varphi}_{m} \partial \hat{\omega}_{m}} & \frac{\partial^2 Q}{\partial \hat{\varphi}_{m}^2}
\end{array}\right)^{-1}\binom{\frac{\partial Q}{\partial \hat{\omega}_{m}}}{\frac{\partial \hat{Q}}{\partial \hat{\varphi}_{m}}},
\end{equation}
for $m=0,1,\cdots,N-1$. Here, $Q\triangleq Q(\hat{\omega}_{m},\hat{\varphi}_{m})$ is an objective function, which is presented as
\begin{equation} \label{norm_gain_v2}
Q(\hat{\omega}_{m}, \hat{\varphi}_{m}) \triangleq \frac{\left|\int_0^{T^{\prime}} \bar{y}(t) \varrho(t) \cos (\hat{\omega}_{m} t+\hat{\varphi}_{m}) \mathrm{d} t\right|^2}{\int_0^{T^{\prime}}|\varrho(t) \cos (\hat{\omega}_{m} t+\hat{\varphi}_{m})|^2 \mathrm{~d} t}.
\end{equation}
Once the updated $Q(\hat{\omega}^{(t)}_{m},\hat{\varphi}^{(t)}_{m})$ is larger than the previous one, the iteration is completed and the final estimated frequency is given by $\hat{\omega}_{m}^{(T_{\mathrm{iter}})}$, where $T_{\mathrm{iter}}$ is the number of total iterations for $m=0,1,\cdots,N-1$.

Eventually, the final estimated frequency can be acquired by taking the average of the estimated frequency, which are presented as $\hat{\omega}^{(A)}=\frac{1}{N}\sum_{m=0}^{N-1}\hat{\omega}^{(T_{\mathrm{iter}})}_{m}$. 
Likewise, the final estimated frequency for frame $B$, $\hat{\omega}^{(B)}=\frac{1}{N}\sum_{m=0}^{N-1}\hat{\omega}^{(T_{\mathrm{iter}})}_{m}$, can be acquired in a same manner by following \eqref{NLS_Prob}-\eqref{norm_gain_v2} for duration $T^{\prime} < t \leq T$.

\subsubsection{\ac{LS}-based Delay-Doppler Estimation} With the estimated fluctuation frequency $\hat{\omega}^{(A)}$ and $\hat{\omega}^{(B)}$, the problem can be formulated based on the \ac{LS} criterion.
Specifically, the fluctuation frequency can be presented as
\begin{equation} \label{LS-equations}
\underbrace{\left[\begin{array}{c}
\hat{\omega}^{(A)} \\
\hat{\omega}^{(B)}
\end{array}\right]}_{\hat{\boldsymbol{\omega}} \in \mathbb{R}^{2 \times 1}}=\underbrace{\left[\begin{array}{cc}
2 \tilde{c}_1^{(A)} & -1 \\
2 \tilde{c}_1^{(B)} & -1
\end{array}\right]}_{\triangleq \mathbf{C}_1 \in \mathbb{R}^{2 \times 2}} \underbrace{\left[\begin{array}{c}
\tau-\tau^{\prime} \\
\nu_r
\end{array}\right]}_{\boldsymbol{\vartheta}},
\end{equation}
where $\hat{\boldsymbol{\omega}}$ is the estimated frequency vector, $\mathbf{C}_{1}$ is the post-chirp matrix for dual-chirp AFDM, and $\boldsymbol{\vartheta}$ is the parameter set of delay-Doppler.
Note that $\mathbf{C}_{1}$ is a full-rank only if $\tilde{c}_1^{(A)}\neq \tilde{c}_1^{(B)}$, which is accomplished by the proposed dual-chirp \ac{AFDM}, and not for the conventional \ac{AFDM}\footnote{Notice that the number of unique chirp rates may be artificially increased (i.e., $c_1^{(C)}, c_1^{(D)}, \ldots$) to obtain an over-determined system to further improve the accuracy of the proposed algorithm, but this trivial extension is not investigated in this article.}.

Finally, applying the \ac{LS} method on \eqref{LS-equations}, we obtain the vector $\boldsymbol{\vartheta}$ as
\begin{equation} \label{eta}
\boldsymbol{\vartheta}=\left(\mathbf{C}_1^{\top} \mathbf{C}_1\right)^{-1} \mathbf{C}_1^{\top}\hat{\boldsymbol{\omega}},
\end{equation}
which yields the estimated delay and Doppler as $\hat{\tau}=[\boldsymbol{\vartheta}]_{1}+\tau^{\prime}$ and $\hat{\nu}_{r}=[\boldsymbol{\vartheta}]_{2}$, where $[\boldsymbol{\vartheta}]_{1}$ and $[\boldsymbol{\vartheta}]_{2}$ are the first and the second elements of $\boldsymbol{\vartheta}$, respectively. 

\section{Simulation Results}
\label{sec:sim_results}
\subsection{Simulation Environment}\label{sec5-1}

\begin{table}[b]
\vspace{-2ex}
\rowcolors{1}{white}{blue!10!white}
\centering
\caption{Parameters for simulations
\label{table: simulation parameters}}
\vspace{-0.5ex}
\resizebox{1\linewidth}{!}{
\begin{tabular}{|c|c|c|}
\hline
\textbf{Description} & \textbf{Parameter} & \textbf{Value}\\ \hline
Energy levels  & $\{|1\rangle, |2\rangle\}$ & $\{6S_{1/2},6P_{3/2}\}$ \\
Energy levels  & $\{|3\rangle, |4\rangle\}$ & $\{60D_{5/2},63P_{3/2}\}$ \\
Frequency & $\omega_{\mathrm{RF}}$ & $2\pi \times 62.76\,{\rm{GHz}}$ \\
Number of chirp-subcarriers & $N$ & 6 \\
Transition dipole moments & $\{\mu_{12},\mu_{34}\}$ & $\{2.586,229\}qa_{0}$ \\
Bohr radius & $a_{0}$ & $52.9\,{\rm{pm}}$ \\
Length of vapor cell & $L$ & $0.02\, \mathrm{m}$ \\
Rabi frequencies & $\{\Omega_{p},\Omega_{c}\}$ & $2\pi \times\{5.8,1\} \, \mathrm{MHz}$ \\
Wavelengths & $\{\lambda_{p},\lambda_{c}\}$ & $\{852,509\}\,{\rm{nm}}$ \\
Decay rate & $\gamma_{2}$ & $2\pi \times 5.2\,\rm{MHz}$ \\
Quantum efficiency & $\eta$ & $0.8$ \\
Target range & $L$ & $1,000\,{\rm{m}}$ \\
Transmitter-to-RARQ distance & $L^{\prime}$ & $1\,{\rm{m}}$ \\
Velocity & $\nu_{r}$ & $50\,{\rm{m/s}}$ \\
Bandwidth & $B$ & $1\,{\rm{MHz}}$ \\
Pre-chirp parameter & $c_{2}$ & $\sqrt{2}$ \\
Density of atoms & $N_{0}$ & $4.89 \times 10^{16}\, \mathrm{m^{-3}}$ \\
\hline
\end{tabular}
}
\vspace{-1.5ex}
\end{table}

Unless stated otherwise, the default setting of the simulations is organized in Table.~\ref{table: simulation parameters}.
The average received \ac{SNR} considering the time-varying amplitude $\varrho_{m}(t)$, is defined as 
\begin{equation} \label{SNR_Definition}
\operatorname{SNR}(t)\triangleq\frac{1}{N}\sum_{m=0}^{N-1}\varrho_{m}^2(t) h^2 T .
\end{equation}
Note that the received \ac{SNR} remains constant when the transmission power is fixed \cite{cui2025realizing}.
For the estimation performance analysis, we adopted the normalized \ac{NRMSE}, which is defined as \cite{ranasinghe2024joint}
\begin{equation} \label{NRMSE}
\mathrm{NRMSE}=\mathbb{E}\left[\frac{1}{|\zeta|}\sqrt{\left(\hat{\zeta}-\zeta\right)^2}\right],
\end{equation}
where $\mathbb{E}[\,\cdot\,]$ is the expectation and  $\zeta$ denotes a given channel parameter (range or velocity) and $\hat{\zeta}$ is its estimate.
%

\subsection{Performance Analysis}\label{sec5-2}

\begin{figure}[t]
\captionsetup[subfigure]{justification=centering}
\begin{center}
\begin{subfigure}[t]{1\columnwidth}
\includegraphics[width=1\columnwidth]{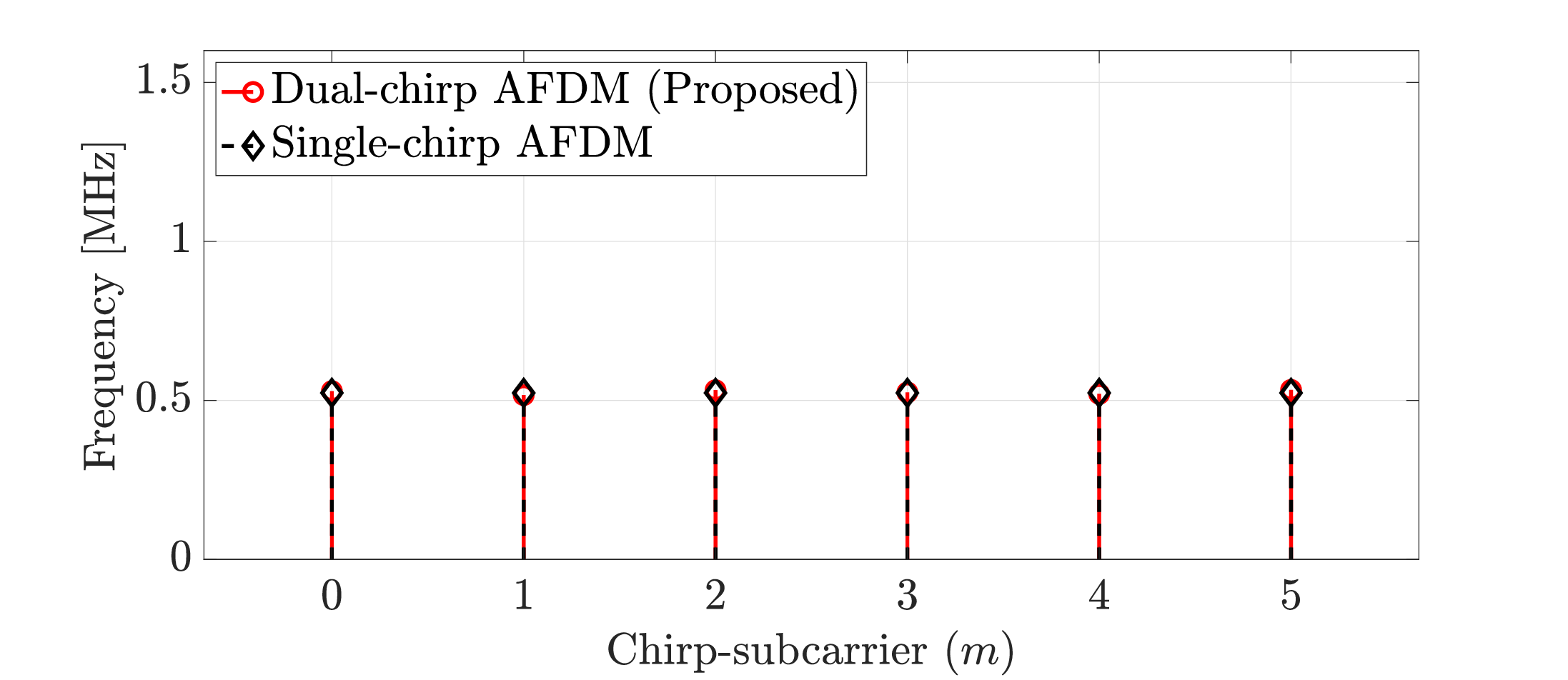}
\caption{\ac{FFT} spectrum of duration $0 < t \leq T^{\prime}$.}
\label{Frame1}
\end{subfigure}
\begin{subfigure}[t]{1\columnwidth}
\includegraphics[width=1\columnwidth]{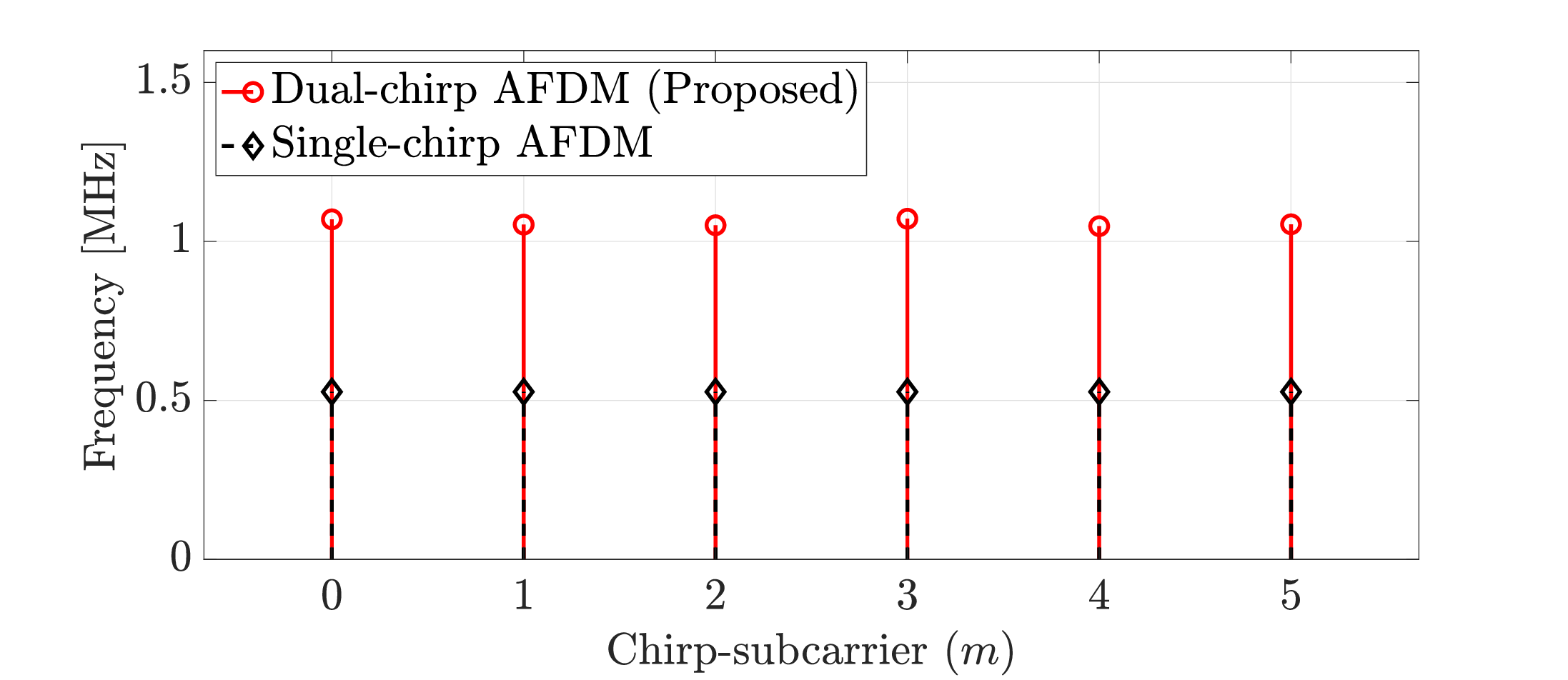}
\caption{\ac{FFT} spectrum of duration $T^{\prime} < t \leq T$.}
\label{Frame2}
\end{subfigure}
\caption{\ac{FFT} spectrum comparison between single-chirp \ac{AFDM} and the proposed dual-chirp \ac{AFDM}, highlighting the different spectra between the two frames of the dual-chirp AFDM, compared to the identical spectra of the conventional AFDM.}\label{FFT-Spectrum} 
\vspace{-2ex}
\end{center}
\end{figure}

In this subsection, we analyze the estimation performance of the proposed dual-chirp \ac{AFDM}.
For the comparison of the delay-Doppler estimation performance, the classical single-chirp \ac{AFDM} is adopted.
To establish a practical performance baseline, the post-chirp difference $\Delta c_{1} = c^{(B)}_{1} - c^{(A)}_{1}$ is introduced to approximate the behavior of conventional single-chirp \ac{AFDM}. 
This formulation is necessary because having exactly identical chirp rates ($c^{(A)}_{1} = c^{(B)}_{1}$), and $\Delta c_1 = 0$ results in a rank-1 measurement matrix $\mathbf{C}_{1}$, which limits the use of \ac{LS}-based delay-Doppler estimation.
By utilizing this approximation, the performance of the conventional framework can be effectively modeled for comparative purposes, while in truth, unambiguous joint delay-Doppler estimation with conventional \ac{AFDM} is not possible.
In the simulation for the single-chirp \ac{AFDM}, we set $\Delta c_{1}$ as $10^{-3}$.

\begin{figure}[t]
\captionsetup[subfigure]{justification=centering}
\begin{center} 
\begin{subfigure}[t]{0.95\columnwidth}
\includegraphics[width=1\columnwidth]{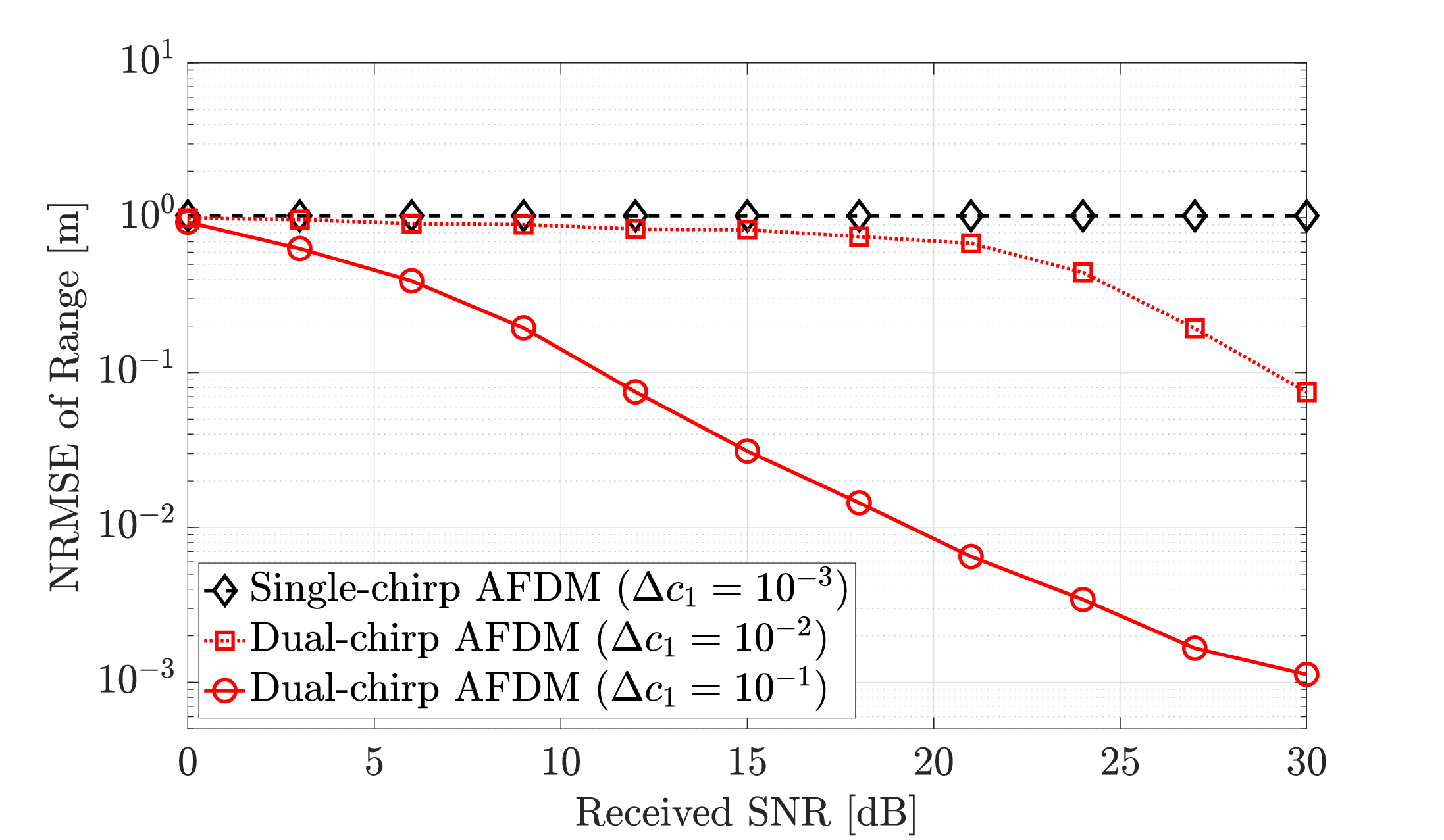}
\caption{\ac{NRMSE} of range.}
\label{NRMSE_Range}
\end{subfigure}
\begin{subfigure}[t]{0.95\columnwidth}
\includegraphics[width=1\columnwidth]{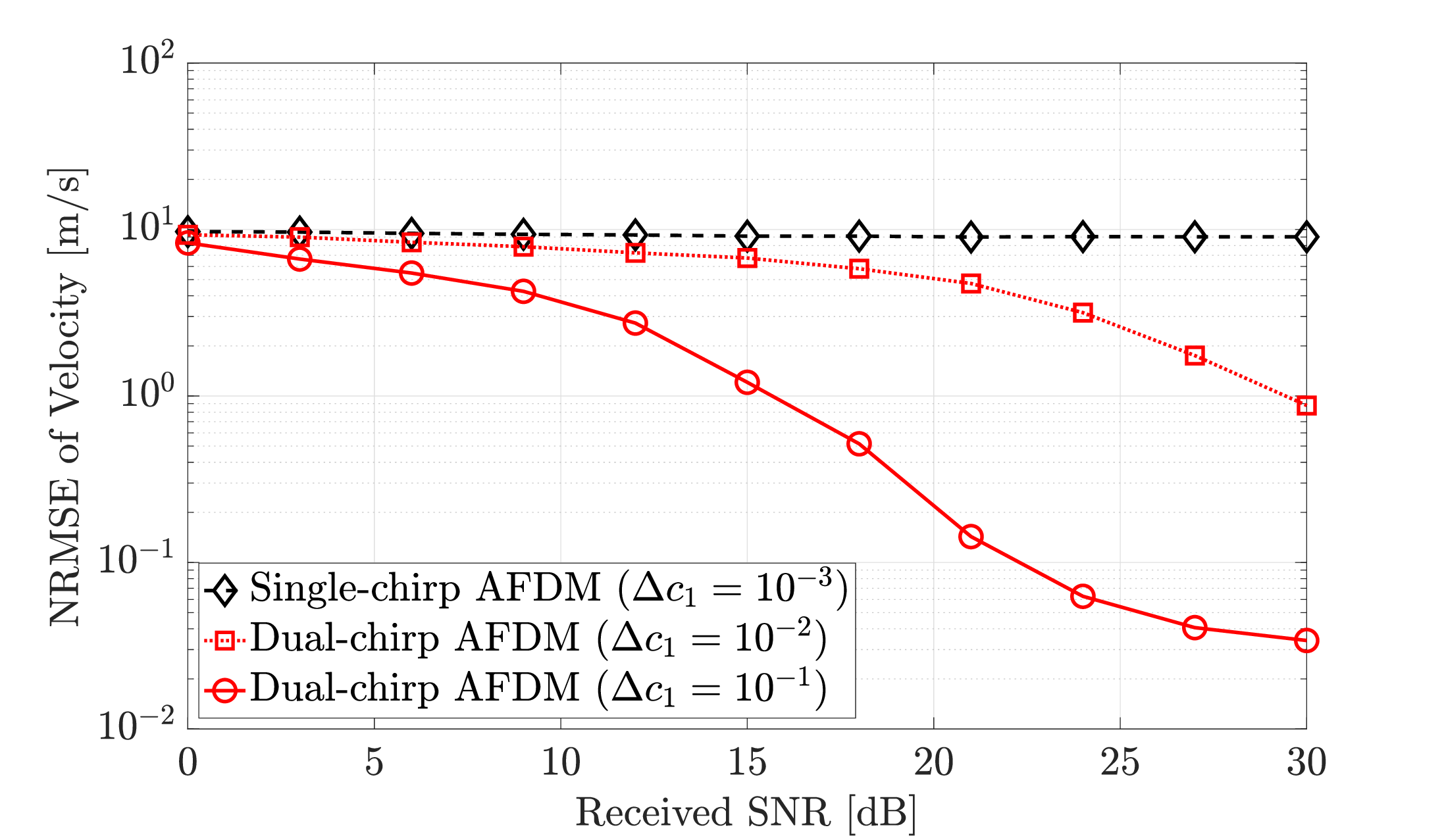}
\caption{\ac{NRMSE} of velocity.}
\label{NRMSE_Velocity}
\end{subfigure}
\caption{Joint range-velocity \ac{NRMSE} performance versus \ac{SNR}.}\label{NRMSE_Analysis} 
\vspace{-4ex}
\end{center}
\end{figure}

First, Fig.~\ref{FFT-Spectrum} presents the results of the \ac{FFT} spectrum analysis, where the average received \ac{SNR} is fixed at 30~dB. 
The post-chirp rates for the dual-chirp \ac{AFDM} are set to $c^{(A)}_{1}=\frac{3}{2\bar{N}}$ and $c^{(B)}_{1}=\frac{6}{2\bar{N}}$, while the single-chirp baseline utilizes $c^{(A)}_{1}=\frac{3}{2\bar{N}}$ and $c^{(B)}_{1}=c^{(A)}_{1}+\Delta c_{1}$. 

As illustrated in the figures, the proposed dual-chirp \ac{AFDM} results in two distinct spectral peaks across the periods $0 < t \leq T^{\prime}$ (Fig.~\ref{Frame1}) and $T^{\prime} < t \leq T$ (Fig.~\ref{Frame2}). 
In contrast, the conventional \ac{AFDM} exhibits nearly identical peaks across both frames, failing to provide additional information for parameter decoupling.
This confirms that the dual-chirp structure effectively generates unique frequency measurements, thereby eliminating the inherent ambiguity in the \ac{RAQR} readout.

Next, Fig.~\ref{NRMSE_Analysis} illustrates the \ac{NRMSE} performance for range and velocity estimation using the proposed approach. 
As expected, the single-chirp \ac{AFDM} baseline (approximated via marginal $\Delta c_1$) exhibits poor estimation accuracy across the entire \ac{SNR} range. 
This error floor stems from the inherent structural ambiguity of the single-chirp waveform, where nearly identical frequency measurements across all subcarriers prevent the decoupling of delay and Doppler parameters.

In contrast, the proposed dual-chirp \ac{AFDM} achieves significantly superior performance.
Furthermore, it is observed that increasing the post-chirp difference $\Delta c_{1}$ leads to an enhancement in estimation accuracy. 
This behavior is attributed to the conditioning of the post-chirp matrix $\mathbf{C}_{1}$, where a negligible $\Delta c_{1}$ makes the matrix near-singular and ill-conditioned, while a sufficient difference restores the full-rank status of $\mathbf{C}_{1}$, thereby enabling robust and unambiguous joint estimation.

\section{Conclusion}
\label{sec:conc}

In this paper, we proposed the dual-chirp \ac{AFDM}, the first waveform tailored for \ac{RAQR}-based joint delay-Doppler estimation. 
By employing distinct post-chirp rates, the proposed framework ensures a full-rank measurement matrix for the delay and Doppler parameters, facilitating accurate parameter recovery via \ac{NLS} and \ac{LS} algorithms.
Numerical results validate the superior performance of dual-chirp \ac{AFDM}, establishing it as a robust solution for emerging quantum sensing and \ac{ISAC} applications.

\bibliographystyle{IEEEtran}
\bibliography{IEEEabrv,reference}
\end{document}

%% file: Acronyms.tex
\acrodef{KPI}[KPI]{key performance indicator}
\acrodef{WRT}[w.r.t.]{with respect to}
\acrodef{WP}[w.p.]{with probability}
\acrodef{LHS}[l.h.s.]{left hand side}
\acrodef{RHS}[r.h.s.]{right hand side}

\acrodef{2D}[2D]{two-dimensional}
\acrodef{3D}[3D]{three-dimensional}
\acrodef{3GPP}[3GPP]{3rd Generation Partnership Project}
\acrodef{AOA}[AOA]{angle-of-arrival}
\acrodef{AWGN}[AWGN]{additive white Gaussian noise}
\acrodef{BS}[BS]{base station}
\acrodef{BFT}[BFT]{beam focusing and tracking}
\acrodef{CFR}[CFR]{channel frequency response}
\acrodef{CIR}[CIR]{channel impulse response}
\acrodef{CTRV}[CTRV]{\textit{Constant Turn Rate and Velocity}}
\acrodef{DDPG}[DDPG]{deep deterministic policy gradient}
\acrodef{DRL}[DRL]{deep reinforcement learning}
\acrodef{ELAA}[ELAA]{extremely large aperture array}
\acrodef{FF}[FF]{far-field}
\acrodef{FR3}[FR3]{frequency range~3}
\acrodef{ISAC}[ISAC]{integrated sensing and communication}
\acrodef{KF}[KF]{Kalman filter}
\acrodef{LMS}[LMS]{least means square}

\acrodef{LOS}[LOS]{line-of-sight}
\acrodef{LS}[LS]{least squares}
\acrodef{LSTM}[LSTM]{long short-term memory}
\acrodef{MAP}[MAP]{maximum a posteriori}
\acrodef{MC}[MC]{Monte Carlo}
\acrodef{MDP}[MDP]{Markov decision process}
\acrodef{MIMO}[MIMO]{multiple-input multiple-output}
\acrodef{ML}[ML]{maximum likelihood}
\acrodef{MLE}[MLE]{maximum likelihood estimation}
\acrodef{MMSE}[MMSE]{minimum mean square error}
\acrodef{MSE}[MSE]{mean square error}
\acrodef{MUI}[MUI]{multi-user interference}
\acrodef{MUSIC}[MUSIC]{multiple signal classification}
\acrodef{NF}[NF]{near-field}
\acrodef{NLOS}[NLOS]{non-line-of-sight}
\acrodef{NN}[NN]{neural network}
\acrodef{OU}[OU]{Ornstein-Uhlenbeck}
\acrodef{PDP}[PDP]{power delay profile}
\acrodef{RCS}[RCS]{radar cross section}
\acrodef{RF}[RF]{radio frequency}
\acrodef{RFID}[RFID]{radio frequency identification}
\acrodef{RFS}[RFS]{random finite set}
\acrodef{RI}[RI]{range information}
\acrodef{RL}[RL]{reinforcement learning}
\acrodef{RMSE}[RMSE]{root-mean-square error}
\acrodef{ReNN}[regression NN]{regression neural network}
\acrodef{ROI}[ROI]{region of interest}
\acrodef{RTT}[RTT]{round-trip time}
\acrodef{RV}[RV]{random variable}
\acrodef{SI}[SI]{soft information}
\acrodef{SIR}[SIR]{signal-to-interference ratio}
\acrodef{SLAM}[SLAM]{simultaneous localization and mapping}
\acrodef{SMC}[SMC]{sequential monte carlo}
\acrodef{SNR}[SNR]{signal-to-noise ratio}
\acrodef{SINR}[SINR]{signal-to-interference-plus-noise ratio}
\acrodef{TDOA}[TDOA]{time difference-of-arrival}
\acrodef{TOA}[TOA]{time-of-arrival}
\acrodef{TOF}[TOF]{time-of-flight}
\acrodef{UAV}[UAV]{unmanned aerial vehicles}
\acrodef{UE}[UE]{user equipment}
\acrodef{UKF}[UKF]{unscented Kalman filter}
\acrodef{UWB}[UWB]{ultra-wideband}
\acrodef{WLS}[WLS]{weighted least squares}
\acrodef{xG}[xG]{Next-generation}
\acrodef{AFDM}[AFDM]{affine frequency division multiplexing}
\acrodef{PD}[PD]{photodetector}
\acrodef{LO}[LO]{local oscillator}
\acrodef{w.r.t.}[w.r.t.]{with respect to}
\acrodef{AFT}[AFT]{affine Fourier transform}
\acrodef{RAQRs}[RAQRs]{Rydberg atomic quantum receivers}
\acrodef{RAQR}[RAQR]{Rydberg atomic quantum receiver}
\acrodef{DAFT}[DAFT]{discrete affine Fourier transform}
\acrodef{NLS}[NLS]{nonlinear least square}
\acrodef{FFT}[FFT]{fast Fourier transform}
\acrodef{ICI}[ICI]{intercarrier interference}
\acrodef{OTFS}[OTFS]{orthogonal time frequency space}
\acrodef{OFDM}[OFDM]{orthogonal time division multiplexing}
\acrodef{DD}[DD]{delay-Doppler}
\acrodef{SQL}[SQL]{standard quantum limit}
\acrodef{IDAFT}[IDAFT]{inverse DAFT}
\acrodef{NRMSE}[NRMSE]{normalized root mean square error}